\documentclass[aps,pra,amssymb,twocolumn,showpacs,supersriptaddress,groupeaddress]{revtex4-1}
\usepackage{amssymb}
\usepackage{amsmath}
\usepackage{graphicx}
\usepackage{dcolumn}
\usepackage{bm}
\usepackage{subcaption}
\usepackage{bm}
\usepackage{natbib,hyperref}
\usepackage{hyperref}
\usepackage{color}
\hypersetup{colorlinks=true, 
    linkcolor=red,          
    citecolor=magenta,        
    filecolor=gree,      
    urlcolor=blue           
}

\usepackage{float}

\begin{document}

\title{Revealing the system-bath coupling via Landau-Zener-St\"uckelberg interferometry in superconducting qubits}

\author{Ana Laura Gramajo$^{1}$, Daniel Dom\'inguez$^{1}$ and Mar\'ia Jos\'e S\'anchez$^{1,2}$}

\affiliation{$^{1}$Centro At\'omico Bariloche and Instituto Balseiro, 8400 San Carlos de Bariloche, R\'io Negro, Argentina.}
\affiliation{$^{2}$Instituto de Nanociencia y Nanotecnolog\'{\i}a (INN),CONICET-CNEA, Argentina.}


\begin{abstract}
In this work we propose a way to unveil the type of environmental noise in strongly driven superconducting flux qubits through the  analysis of  the Landau-Zener-St\"uckelberg (LZS) interferometry. We study both the two-level and the multilevel dynamics of the flux qubit driven by a dc+ac magnetic field. 
We found that  the LZS interference patterns exhibit well defined  multiphoton resonances whose shape strongly depend on the time scale and the type of  coupling to a quantum bath. For the case of transverse system-bath coupling,  the n-photon resonances are narrow and nearly symmetric with respect to the dc magnetic field for almost all time scales, whilst in the case of longitudinal coupling they exhibit a change from a wide symmetric to an antisymmetric  shape for times of the order  of the relaxation time.
We find this dynamic behavior  relevant for the interpretation of several LZS interferometry experiments in which the stationary regime is not completely reached.
  
\end{abstract}


\maketitle


Superconducting circuits with  Josephson junctions \cite{orlando_1999,chiorescu_2003}  behave  as artificial 
atoms \cite{you_2011} and have been  extensively proven as quantum bits \cite{makhlin_2001}.
When driven by  a dc+ac magnetic flux, Landau-Zener-St\"uckelberg
(LZS) interference patterns \cite{shevchenko_2010} combined with multi-photon resonances 
have been  observed \cite{oliver_2005,berns_2006,berns_2008,oliver_2009,izmalkov_2008} 
and  used to probe the energy level spectrum of the device for large driving amplitudes. \cite{berns_2008,oliver_2009}.
LZS patterns  also emerge  in
charge qubits\cite{sillanpaa_2006,wilson_2007}, Rydberg atoms\cite{yoakum_1992}, ultracold molecular gases\cite{mark_2007},
optical lattices\cite{kling_2010} and single electron spins systems\cite{huang_2011}.
In addition, LZS interferometry  was recently proposed as a tool to determine
relevant information related to the coupling of a  qubit with a noisy environment, such as 
dissipation strength and dephasing time \cite{forster_2014,blattmann_2015,mi_2018}.
These studies have been performed for steady state-experiments, where full relaxation with the bath degrees of freedom is assumed.

In the present work we demonstrate that the finite time LZS spectroscopy can unveil additional features linked to how relevant times scales affect the symmetry of the resonance patterns for  different system-bath couplings.
As a system  of study we chose  the superconducting Flux Qubit (FQ) originally introduced in Ref.[\onlinecite{orlando_1999}] and
which over the last years, due to the improvement in its  design and fabrication techniques, has  become one of  the most tested devices for  quantum information proposals \cite{fei_2016}.  Recent experiments on the FQ have implemented
noise spectroscopy for  different sources of noise (flux noise, charge noise, critical current noise) through dynamical decoupling  \cite{bylander_2011} and driven evolution measurements \cite{yan_2013,yoshihara_2014}. 

Here to address  
the finite time LZS interferometry  we  study the FQ coupled to a quantum bath and under strong periodic driving, using the Floquet-Markov quantum master equations \cite{grifoni_1998,kohler_1998}.
Our main finding is that a dynamic change in the symmetry of a  n-photon resonance 
takes place for  the case of longitudinal system-bath coupling, whilst  the resonances 
remain almost undisturbed in time for  transverse system-bath coupling. 

Our analysis becomes particularly relevant to understand LZS interferometry experiments for  FQ with large relaxation times \cite{oliver_2005,oliver_2009}. Several well established theoretical works have studied the steady state of periodically driven two level systems \cite{hartmann_2000,dakhnovskii_1995,stace_2005,goorden_2004,hausinger_2010}. However, the experimental results on LZS interferometry in the FQ  do not agree with these previous theoretical results. The theory of   \cite{hartmann_2000,dakhnovskii_1995,stace_2005,goorden_2004,hausinger_2010} shows population inversion and antisymmetric resonance patterns as a function of the energy detuning, instead of the symmetric patterns observed in the FQ experiments \cite{oliver_2005,oliver_2009}. A possible explanation was put forward in Refs.\cite{ferron_2012,ferron_2016}: there is a dynamic transition from symmetric resonance patterns below the relaxation time $t_r$ to antisymmetric resonance patterns for time scales above $t_r$. Since the FQ experiments were performed at finite times scales below $t_r$, the steady state patterns were not observed, according to this scenario. On the other hand, in Ref.[\onlinecite{blattmann_2015}] it was shown that  transverse noise (previous works \cite{hartmann_2000,dakhnovskii_1995,stace_2005,goorden_2004,hausinger_2010,ferron_2012,ferron_2016} considered longitudinal noise) can lead to steady state symmetric resonances in LZS interferometry, which suggest a different possible explanation of the experimental results.
The aim of this work is to assess which scenario is more adequate to explain the experiments of Refs.\cite{oliver_2005,oliver_2009} by analyzing the time dependence of the LZS  patterns for different system-bath couplings (transverse and longitudinal noise).

We start in Sec. I by writing the  Hamiltonian of  the FQ in the presence of different sources of  quantum noise and describing the Floquet-Markov formulation for open quantum systems with a time periodic drive.
In Sec. II we show results for the time dependent evolution of the driven FQ with different sources of noise, restricted to a two levels system (TLS) regime.
In Sec. III we extend the analysis to the multilevel case, which is relevant for large driving amplitudes and  to compare with LZS experiments. Conclusions are given in Sec. IV.

\section{Dynamics of the Flux Qubit}
\label{sec:I}

\subsection{The Flux Qubit and noise sources}
\label{subsec:IA}
The FQ consists on a   superconducting ring with three Josephson
junctions\cite{orlando_1999} enclosing a magnetic flux $\Phi=
f\Phi_0$ ($\Phi_0=h/2e$) with
phase differences
$\varphi_1$, $\varphi_2$ and
$\varphi_3=-\varphi _1 +\varphi _2 -2\pi f$.
Two of the junctions have  coupling energy,
$E_J$, and capacitance, $C$,
while the third
has  $E_{J,3}=\alpha E_J$ and $C_3=\alpha C$. 
In the quantum regime, the FQ  Hamiltonian reads:\cite{orlando_1999}
\begin{equation}
\label{eq:1}
{{\cal H}}_{FQ}= E_p { n}_p^2 + E_m { n}_m^2 +E_J{V}\; ,
\end{equation}
with  $\varphi_p=\frac{\varphi_1+\varphi_2}{2}$ and 
$\varphi_m=\frac{\varphi_1-\varphi_2}{2}$ the phase operators,
 ${ n}_{k}=-i\frac{\partial}{\partial  \varphi_{k}}$ ($k=p,m$) the
 charge number operators,  $E_{p}=2E_{C}$, $E_{m}=\frac{E_{p}}{1+2\alpha}$, 
$E_C=e^2/2C$ and ${V}(\varphi_p,\varphi_m;f)= 2+\alpha -
2\cos\varphi_p\cos \varphi_m - \alpha \cos (2\pi f+2\varphi _m)$.
The FQ has several levels with eigenenergies $E_i$ 
and eigenstates  $|\Psi_i\rangle$
which depend on $\alpha$,  $\eta=\sqrt{8E_C/E_J}$ and  flux detuning ${\tilde f}=f-1/2$.
Typical experiments  have $\alpha \sim 0.6-0.9$ and $\eta\sim 0.1-0.6$ 
\cite{chiorescu_2003,oliver_2005,berns_2006,berns_2008,oliver_2009}.
For $\alpha \ge 1/2$ and
$|{\tilde f}| \ll 1$, the potential ${V}$
has the shape of a double-well with two minima along the $\varphi_l$ direction.
Each minima corresponds
to macroscopic persistent currents of opposite sign, and 
for ${\tilde f}\gtrsim 0$ (${\tilde f} \lesssim 0$) a ground state with positive (negative) loop current is
favored.
 In this regime the system can be operated as a
quantum bit\cite{orlando_1999,chiorescu_2003} and  approximated
by a two-level system (TLS)\cite{orlando_1999,ferron_2010}. 

The main sources of relaxation and decoherence in
the FQ are flux noise $\delta f(t)$, charge noise
$\delta  N(t)$, and critical current noise $\delta {I}_c(t)$ \cite{you_2007,yan_2013,yoshihara_2014,bylander_2011,sete_2017}.
In the case of weak fluctuations, the different sources of  noise
can be incorporated in Eq.(\ref{eq:1}) by the replacements $f\rightarrow f+{\delta f}$,
${ n}_k\rightarrow { n}_k-{\delta  N}_k$ ($k=p,m$),
and $E_J \rightarrow E_J(1+{\delta  I_c}/{I}_c)$, respectively  \cite{you_2007,yan_2013,yoshihara_2014,bylander_2011}.
This leads to ${\cal H}_{FQ}\rightarrow{\cal H}_{FQ}'\approx {\cal H}_{FQ}+{H}_{int}$, where
\begin{equation}
	{H}_{int}={H}_{int}^{ch} + {H}_{int}^{f}+{H}_{int}^{cc}\;,
	\label{eq:2}
\end{equation}
and
\begin{eqnarray}
{H}_{int}^{ch}&=&-2E_{p}{ n}_{p}\delta{N}_{p}
- 2E_{m}{ n}_{m}\delta{N}_{m}\;,\nonumber\\
{H}_{int}^{f}&=&-{2\pi E_{J}}{ I}\delta{f}\;,\nonumber\\
{H}_{int}^{cc}&=&\frac{E_J}{I_c}{ V}\delta{ I_c}\;,
\label{eq:3}
\end{eqnarray} 
with ${ I}=\alpha \sin (2\pi {\tilde f} + 2{\varphi}_{m})$,
the loop current operator normalized by $I_{c}=\frac{2\pi E_{J}}{\Phi_{0}}$.
Notice that Eq.(\ref{eq:2}) results from neglecting quadratic terms in $({ n}_{p}- {N}_{p})^{2}$ and $({ n}_{m}-{N}_m)^{2}$, since we are assuming  the weak fluctuations regime.

If we consider the lowest eigenstates,
the  term with ${ n}_{p}$  can be neglected\cite{foot1}
and we can redefine the system-bath interaction Hamiltonian as
\begin{equation}
{H}_{int}={\cal A}^{ch}\otimes{\cal B}^{ch}+{\cal A}^{f}\otimes{\cal B}^{f}+{\cal A}^{cc}\otimes{\cal B}^{cc},
\label{eq:4}
\end{equation} 
where
the  system operators are ${\cal A}^{ch}=- 2E_{m}{ n}_{m}$, ${\cal A}^{f}=-{2\pi E_{J}} { I}$,  ${\cal A}^{cc}=E_{J}{ V}$;  and the 
 normalized bath (noise) operators are
${\cal B}^{ch}=\delta{N}_{p}$, ${\cal B}^{f}=\delta{f}$ and
${\cal B}^{cc}=\delta{ I_c}/I_c$.

As a first approach we will consider in Sec. II the FQ restricted to the
two-lowest computational levels, \cite{orlando_1999,ferron_2010} 
\begin{equation}
{\cal H}_{TLS}=-\frac{\epsilon}{2} {\sigma}_z - \frac{\Delta}{2} {\sigma}_x
\;,
\label{eq:5}
\end{equation}
where the Hamiltonian is written 
in the basis defined by the persistent current states 
$|+\rangle=(|0\rangle+|1\rangle)/\sqrt{2}$ and $|-\rangle=(|0\rangle-|1\rangle)/\sqrt{2}$, where $|0\rangle$ and $|1\rangle$
are the ground and excited  FQ states at $\delta f=0$.
The parameters of ${\cal H}_{TLS}$ are the detuning 
$\epsilon= 4\pi I_p {\tilde f}$, and  the energy gap  $\Delta=E_1-E_0$ at ${\tilde f}=0$.
Here $I_p= |\langle+|{ I}|+\rangle|=|\langle-|{ I}|-\rangle|$ is the magnitude of the loop current.
Within this approximation, the noise coupling operators become
\begin{eqnarray}
{\mathcal{A}}^{f}&=&-\lambda_f{\sigma}_{z},\nonumber  \\
{\mathcal{A}}^{ch}&=&-\lambda_{ch}{\sigma}_{y},\nonumber\\
{\mathcal{A}}^{cc}&=&-\lambda_{cc}{\sigma}_{x},	\, ,
\label{eq:6}
\end{eqnarray} 
with $\lambda_f=2\pi |\langle+|{ I}|+\rangle|$, $\lambda_{ch}=\frac{\eta^2}{4+8\alpha}|\langle-|{ n}_m|+\rangle|$ and
$\lambda_{cc}=-|\langle-|{ V}|+\rangle|$.

For the parameter values $\alpha=0.8$ and $\eta=0.25$ and after diagonalization of ${\cal H}_{TLS}$, we obtain  $I_p=0.721$ (in units of $ E_J/\Phi_0$)
and $\Delta=3.33\times 10^{-4}$ (in units of $E_J$). Thus, the noise 
coupling parameters results
$\lambda_f\approx4.5$,
$\lambda_{ch}\approx 3\times 10^{-4}$,
 and
$\lambda_{cc}\approx 4\times 10^{-3}$ (the neglected term corresponding to ${ n}_{p}$ has coupling parameter $\lambda^{p}_{ch}=\frac{\eta^2}{4}|\langle-|{ n}_p|+\rangle|\approx  10^{-13}$).
%
%
%
%
%
%
%
%
%


\subsection{LZS interferometry in the presence of quantum noise: The Floquet-Markov approach}
\label{subsec:IC}

In experiments with flux qubits, LZS interferometry \cite{oliver_2005,berns_2006,berns_2008,oliver_2009,rudner_2008} is performed  applying an  harmonic (ac) field of frequency $\omega_{0}$ on top of the static flux, i.e.
\begin{equation}
{\tilde f}\rightarrow  {\tilde f}(t)= {\tilde f}_{dc}+f_{ac}\cos{(\omega_{0} t)} \;.
\label{eq:7}
\end{equation}

In this work, and following Refs.[\onlinecite{shirley_1965,grifoni_1998,breuer_2000,hone_2009,hausinger_2010,ferron2_2010,ferron_2012,ferron_2016}] we analyze  the  LZS interferometry employing  the Floquet formalism, which allows for an 
exact treatment of  harmonic drivings  of arbitrary strength and frequency.
Alternative  approaches to the description of the LZS interference patterns rely on approximations valid either for large driving frequencies or  low driving amplitudes \cite{berns_2006,ashhab_2007, shevchenko_2010}.

For the harmonic driving, the Hamiltonian of the FQ results 
time periodic ${\cal H}_{FQ}(t) = {\cal H}_{FQ}(t + \tau)$, with $\tau=2\pi/\omega_0$.
In  the Floquet formalism, the solutions of the    Schr\"odinger equation are of the
form  $|\Psi_\alpha(t)\rangle=e^{i\varepsilon_\alpha t/\hbar}|\alpha(t)\rangle$, where
the  Floquet states $|\alpha(t)\rangle$
satisfy $|\alpha(t)\rangle$=$|\alpha(t+ \tau)\rangle = 
\sum_k |\alpha_{k} \rangle e^{-ik\omega t}$, and
are eigenstates of the equation
$[{\cal H} (t)- i \hbar \partial/\partial t ] |\alpha(t)\rangle= \varepsilon_\alpha |\alpha(t)\rangle$,
with $\varepsilon_\alpha$ the associated quasi-energy.

Since the FQ is in contact with the environment,
the total Hamiltonian  of the open system is 
$$ H= H_s(t) +  H_B + {H}_{int} .$$ 
Here, $H_s$ is the system Hamiltonian, 
in our case $ H_s={\cal H}_{FQ}$, 
$ H_B$ is the Hamiltonian of the environment, which is
usually modeled as a bath 
of harmonic oscillators\cite{kohler_1998,breuer_2000,hone_2009,hausinger_2010,goorden_2004,goorden_2005,vanderWal_2003}, and
$ {H}_{int}$  is the system-bath interaction Hamiltonian. 
For  weak coupling  (Born approximation) 
and fast bath relaxation (Markov approximation),
a Floquet-Born-Markov master  equation  
for   the system reduced density matrix $\rho$
in the Floquet basis,
$\rho_{\alpha\beta}(t)=\langle\alpha(t)|\rho(t)|\beta(t)\rangle$,
can be obtained\cite{kohler_1998,breuer_2000,hone_2009}:
\begin{eqnarray}
\frac{d\rho_{\alpha\beta}(t)}{dt}&=&
\sum_{\alpha'\beta'} \Lambda_{\alpha\beta\alpha'\beta'}\;\rho_{\alpha'\beta'}\,,
\nonumber\\
\Lambda_{\alpha\beta\alpha'\beta'}&=&-\frac{i}{\hbar}
(\varepsilon_\alpha-\varepsilon_\beta)\delta_{\alpha\alpha'}\delta_{\beta\beta'}+ L_{\alpha\beta\alpha'\beta'}\;.
\label{eq:8}
\end{eqnarray}
The coefficients  ${\it L}_{\alpha\beta\alpha'\beta'}$ are usually rewritten in terms of transition rates $R_{\alpha\beta\alpha'\beta'}$ as
\begin{eqnarray}
{\it L}_{\alpha\beta\alpha'\beta'}&=& R_{\alpha\beta\alpha'\beta'}
+R_{\beta\alpha\beta'\alpha'}^* \\
& &-\sum_\eta \left( \delta_{\beta \beta'}
R_{\eta\eta\alpha'\alpha}+ 
\delta_{\alpha \alpha'}R_{\eta\eta\beta'\beta}^* \right).
\nonumber
\label{eq:9}
\end{eqnarray} 

As we already shown in Eq.(\ref{eq:4}), the interaction Hamiltonian can be written as
$${H}_{int}=\sum_{\nu} {\cal A}^{\nu}\otimes{\cal B}^{\nu},$$
where the ${\cal A}^{\nu}$ are system operators and the ${\cal B}^{\nu}$
are bath operators associated  to different noise sources.
In the case of independent noise sources, 
the corresponding bath operators are  uncorrelated such that $\langle{\cal B^{\nu}}(t){\cal B}^{\nu'}(t')  \rangle=0$ for $\nu\not=\nu'$, and 
the transition rates $R_{\alpha\beta\alpha'\beta'}$ are given as
$$R_{\alpha\beta\alpha'\beta'}=\sum_{\nu} R_{\alpha\beta\alpha'\beta'}^{\nu},$$
with 
\begin{equation}
R_{\alpha\beta\alpha'\beta'}^{\nu} = \sum_{q} g^{\nu}(\omega_{\alpha \alpha',q}) A_{\alpha \alpha',q}^{\nu}A_{\beta' \beta,-q}^{\nu}\,,
\label{eq:10}
\end{equation}
where $\omega_{\alpha \alpha',q}=
(\varepsilon_\alpha-\varepsilon_\alpha')/\hbar+q\omega_{0}$.
In this way, the system-bath interaction is encoded
in the transition matrix elements
$$A_{\alpha\beta,q}^{\nu}=\sum_{k}\langle\alpha_{k}| {\cal A}^{\nu}  |\beta_{k+q}\rangle.$$
Assuming that each bath is in equilibrium  at temperature $T^{\nu}$ it is customary to define: 
$$g^{\nu}({\omega})=\int_{-\infty}^{\infty}dt\langle{\cal B^{\nu}}(t){\cal B^{\nu}}(0)  \rangle  e^{-i\omega t}=J^{\nu}(\omega)n_{\rm th}^{\nu}(\hbar\omega),$$ 
where   $J^{\nu}(\omega)$ [defining $J^{\nu}(-\omega)=-J^{\nu}(\omega)$] is the bath spectral density  and
$n_{\rm th}^{\nu}(\varepsilon)=1/[\exp{(\varepsilon/k_BT^{\nu})}-1]$  \cite{comment4}.


In this work, we compute numerically the  Floquet states $|\alpha(t)\rangle$,
and calculate the coefficients  ${\it L}_{\alpha\beta\alpha'\beta'}$ ,
from which the solution of $\rho_{\alpha\beta}(t)$ can be obtained \cite{ferron2_2010,ferron_2016}.

\section{Two levels regime}
\label{sec:II}
\subsection{Unitary evolution and  LZS interferometry}
\label{subsec:IB}
We start by reviewing the LZS interferometry for the TLS.
In order to obtain the driven  Hamiltonian,  we replace in  ${\cal H}_{TLS}$ 
\begin{equation}
\epsilon\rightarrow \epsilon (t)= \epsilon_{0} + A \cos(\omega_{0} t),
\label{eq:11}
\end{equation}
with $\epsilon_{0}=4\pi I_p {\tilde f}_{dc}$ and $A = 4\pi I_p  f_{ac}$.
The frequency $\omega_0$ is written in units of $E_J/\hbar$ and the qubit eigenenergies in 
units of $E_J$.
When $f_{ac}>|{\tilde f}_{dc}|$ the central avoided crossing at ${\tilde f}=0$ 
is reached for driving amplitudes, ${\tilde f}_{dc} \pm f_{ac}$. In this case
the periodically repeated transitions at ${\tilde f}=0$ give rise  to the LZS 
interference patterns as a function of ${\tilde f}_{dc}$  and $f_{ac}$, characterized
by {\it multiphoton resonances} at $E_1({\tilde f})-E_0({\tilde f})= n\omega_{0}$ \cite{shevchenko_2010,grifoni_1998,breuer_2000,hone_2009,hausinger_2010,ashhab_2007} .

For  $\epsilon_0\gg \Delta$ the resonances take place at $\epsilon_0\simeq n\omega_0$,
and denoting $f_\omega = \omega_0/4\pi I_p$, the $n$-resonance condition can  be written as $\epsilon_0/\omega_0={\tilde f} _{dc}/f_\omega=n$.

In the regime $A\omega_{0} \gg \Delta^{2}$ and  in the Rotating Wave Approximation (RWA) \cite{shevchenko_2010,ashhab_2007,berns_2006},  
the time averaged probability of measuring a positive loop current 
$\overline{ P_+}=\overline{|\langle\Psi(t)|+\rangle|^2}$,  near a $n-$ photon resonance results
\begin{equation}
\overline{P_{+}}=1-\frac{1}{2} \frac{\Delta_{n}^{2}}
{(n \omega_{0} - {\epsilon}_{0})^{2} + \Delta_{n}^{2}}  \; . 
\label{eq:12}
\end{equation}
When the resonance condition  $\epsilon_0=n\omega_0$ is satisfied,  Eq.(\ref{eq:12})
gives $\overline{P_{+}}=1/2$, otherwise is $\overline{P_{+}}\lesssim 1$.
Furthermore, the width of the resonance is  $\delta\epsilon=|\Delta_n|=\Delta |J_{n} (A/ \omega_{0})|=\Delta |J_{n} (f_{ac}/f_\omega)|$, with  $J_n(x)$ the Bessel function of  first kind. 
This gives a quasi periodic dependence as a function of $f_{ac}$ for ${\tilde f}_{dc}$ fixed
near the resonance. In particular, at the zeros of $J_n(x)$
the resonance is destroyed, with $\overline{ P_+}\sim 1$ instead of $\overline{ P_+}=1/2$, a 
phenomenon known as {\it coherent destruction of tunneling} \cite{grossmann_1991,kayanuma_2008}. 
Plots of $\overline{P_{+}}$ as a function of flux detuning ${\tilde f}_{dc}$ and ac amplitude $f_{ac}$ give the typical 
LZS interference patterns, which have been measured experimentally  in flux qubits \cite{oliver_2005,berns_2006,berns_2008,oliver_2009,rudner_2008} and have also been observed in other driven systems \cite{izmalkov_2008,sillanpaa_2006,wilson_2007,wilson_2010,sun_2009,sun_2011,wang_2010,degraaf_2013,shevchenko_2012,mark_2007,petta_2010,stehlik_2012,cao_2013,dupont_2013,shang_2013,nalbach_2013,forster_2014,granger_2015,huang_2011,zhou_2014,lahaye_2009,kling_2010}.

Several phenomenological approaches \cite{shevchenko_2010,berns_2008,du_2010,you_2011}
have taken into account relaxation and decoherence effects in LZS interferometry, obtaining a broadening of the Lorentzian-shape $n-$ photon resonances of Eq.(\ref{eq:12}). 

\subsection{Longitudinal vs. transverse noise}
\label{subsec:IIA}

In this section we analyze the environmental  noise employing  the Floquet Markov master equation, described in Sec.\ref{subsec:IC}.
We first consider the two extreme cases: either  pure
flux or  ``longitudinal'' noise (which commutes with the driving),
or  pure charge noise, which we call ``transverse'' noise.
For simplicity, we  consider in both cases that the 
baths are equilibrated at the same temperature $T^\nu=T$ with an ohmic spectral  density $J^\nu(\omega)=\gamma\omega e^{-\omega/\omega_c}$.
In the case of pure longitudinal noise we consider, ${\mathcal{A}}^{f}\rightarrow{\mathcal{A}}=- \lambda_f {\sigma}_z$
while in the
case of pure transverse noise we take  ${\mathcal{A}}^{ch}\rightarrow{\mathcal{A}}=-\lambda_{ch}{\sigma}_y$.
In order to establish a quantitative comparison among the two types of noise we  first analize the results for equal coupling strengths $\lambda^{(f)}=\lambda^{(ch)}=1$. 
 
 We use typical reported experimental values for FQ \cite{oliver_2005}, $E_J/h \sim 300{\rm GHz}$, driving  microwave frequency $\omega_{0}/2\pi=0.003E_J/\hbar\sim 900\rm{MHz}$,  bath temperature  $T=0.0014 E_J/k_B (\sim 20{\rm mK})$ and
 we consider $\gamma=0.001$.
Furthermore, in all the cases we are assuming that the FQ is initially prepared in its ground state  $|\Psi_0\rangle$ of the static Hamiltonian $H_0\equiv{\cal H}_{FQ}({\tilde f}={\tilde f}_{dc})$.
Experimentally, the probability of having a state of positive or negative
persistent current in the FQ  is measured\cite{chiorescu_2003,oliver_2005}. The probability
of a positive current measurement  can
be calculated as
$P_+(t)={\rm Tr}(\Pi_+  \rho(t))$, with $\Pi_+=|+\rangle\langle+|$. 
For a static detuning   ${\tilde f}_{dc}\gtrsim 0$,  the ground state has $P_+ (0) \approx 1$.
\begin{figure*}[htb!]	
	\centering
		\includegraphics[width=0.9\linewidth]{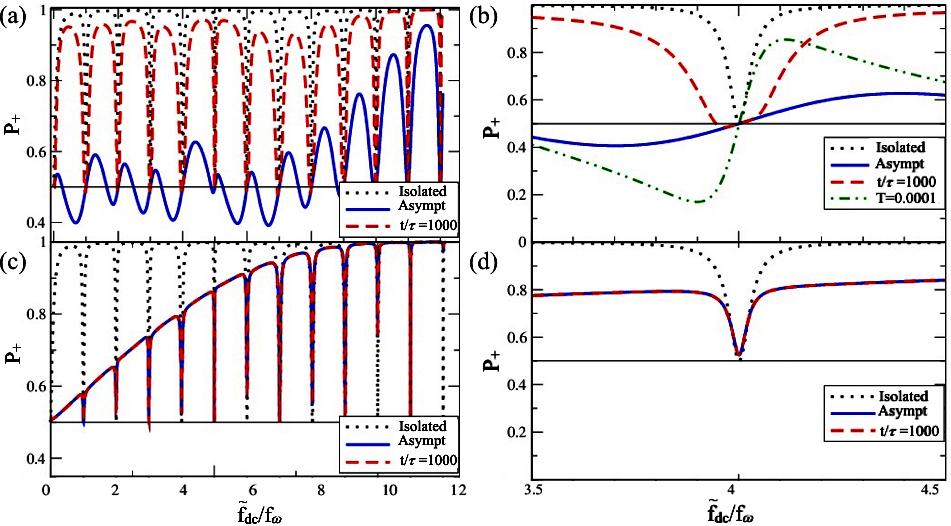}	
	\caption{Population $P_+$ as a function of the $dc$ flux detuning ${\tilde f}_{dc}$ (normalized by $f_\omega=\omega_0/4\pi I_p$) for the FQ  restricted to TLS, driven with amplitude	$f_{ac}=0.003$ and $\omega_{0}= 2\pi/\tau= 0.003 E_J/\hbar$. The Ohmic bath is at $T=0.0014$ ($20$mK for $E_J/h\approx300{\rm GHz}$). Dotted line: $\overline P_+$ for the isolated system; solid line:  $P_+ (t=1000 \tau)$, dashed line: asymptotic ($t\rightarrow\infty$) $\overline P_+$.
	Horizontal  solid line indicates $P_+=0.5$ value. (a) Longitudinal coupling for an Ohmic bath with $\gamma^{(f)}=0.001$ and (b) enlarged view of (a)  around n=4 photon resonance. (c) Transverse coupling for an Ohmic bath with $\gamma^{(ch)}=0.001$ and (d) enlarged view of (c) around n=4 photon resonance.} 
	\label{fig:1}
\end{figure*}
\begin{figure*}[htb!]
	\centering
	\includegraphics[width=0.9\linewidth]{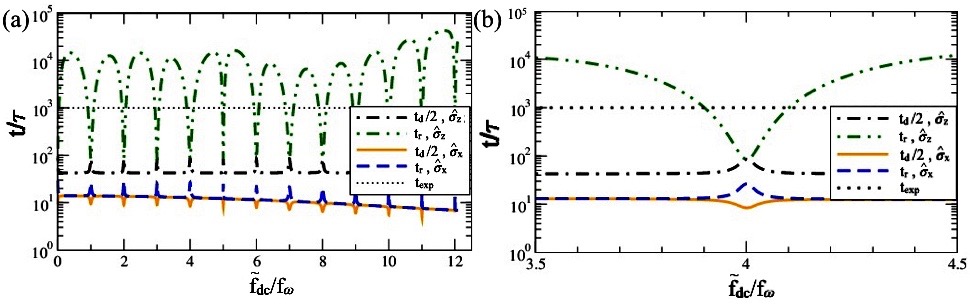}	
	\caption{(a) Relaxation time $ t_r$ for the longitudinal (dashed-double dotted line) and transverse (dashed line) couplings. (Half) decoherence time $t_{d} /2$ for the longitudinal (dashed-dotted line) and transverse (solid line) couplings. The flux detuning is normalized by $f_\omega=\omega_0/4\pi I_p$, such that the $n$-photon resonances are at ${\tilde f}=nf_\omega$. (b) Enlarged view of (a) around the $n=4$ resonance. The experimental time $t_{exp}/\tau=$ 1000 is plotted by the dotted line.} 
	\label{fig:2}
\end{figure*}

In Figs.\ref{fig:1}(a-b) and Figs.\ref{fig:1}(c-d) we plot $P_+$
for  longitudinal and transverse couplings respectively, as a function
of the flux detuning ${\tilde f}_{dc}$ for a fixed value of $f_{ac}=0.003$. 
As a comparison, for both couplings we  plot  $\overline P_+$  for the isolated case (without dissipation), where the $n$-photon resonances are clearly displayed as minima at  $\epsilon_0/\omega_0={\tilde f} _{dc}/f_\omega=n$.  
For longitudinal coupling,  Fig.\ref{fig:1}(a)  shows that
for  time scales of  FQ  experiments \cite{oliver_2005} (we take here as a typical value $t_{exp}=1000\tau$), the 
behavior of  $P_+$  is similar to the isolated case, with a broadening of the
minima at the multiphoton resonances due to decoherence.
On the other hand, in the asymptotic $t\rightarrow\infty$ steady state,
  $P_+$ exhibits  antisymmetric
multiphoton resonances \cite{blattmann_2015,ferron_2016}, clearly displayed in 
 Fig.\ref{fig:1}(b), where an enlarged view  around the $n=4$ resonance is shown. 
 Morover, as the temperature is lowered, the antisymmetry around the resonance condition is more evident, as it is shown for $T=0.0001 E_{J}/k_{B}$. 

For transverse coupling, see Figs.\ref{fig:1}(c) and (d), the behavior of  $P_+$ vs. ${\tilde f}_{dc}$ is remarkably
different from the previous case: (i) there are no noticeable differences between
the finite time and the steady sate $P_+$; (ii) the multiphoton resonances are symmetric in the 
steady state; (iii) there is no broadening of the resonances compared 
to the isolated case; and (iv) there is a linear background in  $P_+$ as a function 
of ${\tilde f}_{dc}$ for the off-resonant situations.  The hallmarks (ii) and (iv) have
been also found in Ref.[\onlinecite{blattmann_2015}].
The linear background in  $P_+$  can be understood by a simple argument. The transverse coupling through $\sigma_y$   provides a direct  relaxation mechanism to the ground state (same holds for $\sigma_x$ coupling). Assuming that for the off-resonant situations in the steady state the qubit is fully relaxed in the ground state,  we can estimate that in average is $P_{+}\sim t'/\tau$,  with
$t'$ the time scale within one period $\tau$  in which the ground state has nearly complete overlap with the $|+\rangle$ state (when ${\tilde f}(t) > 0$).
For small ${\tilde f}_{dc}/f_{ac}$ this gives $P_{+} \sim \frac{1}{2} + \frac{1}{\pi}  \frac{ {\tilde f}_{dc}}{ f_{ac}}$. This straightforward calculation illustrates the linear background in the  dependence  of $P_{+}$ with $f_{dc}$ observed in Fig\ref{fig:1}(c).

The above described features have its correlation in the behaviour  of   the relaxation ($t_r$) and the decoherence ($t_d$) times, which are shown in Fig.\ref{fig:2} for both types of couplings. They are calculated numerically from the eigenvalues of ${\Lambda}$ defined in Eq.(\ref{eq:10}),
the maximum non-zero real eigenvalue of ${\Lambda}$ gives $-t_r^{-1}$, and the
real part of the complex conjugates eigenvalues of ${\Lambda}$ give $-t_d^{-1}$ \cite{ferron2_2010,ferron_2016}.
In general is  $\frac{1}{t_d}=\frac{1}{2t_r}+\frac{1}{t_\phi}$ with $t_\phi$ the dephasing time and thus  the decoherence time satisfies $t_d\le 2t_r$ \cite{grifoni_1998}. 

\begin{figure}[htb!]
	\begin{subfigure}[ht!]{0.45\textwidth} 
	\centering
		\includegraphics[width=0.85\linewidth,clip]{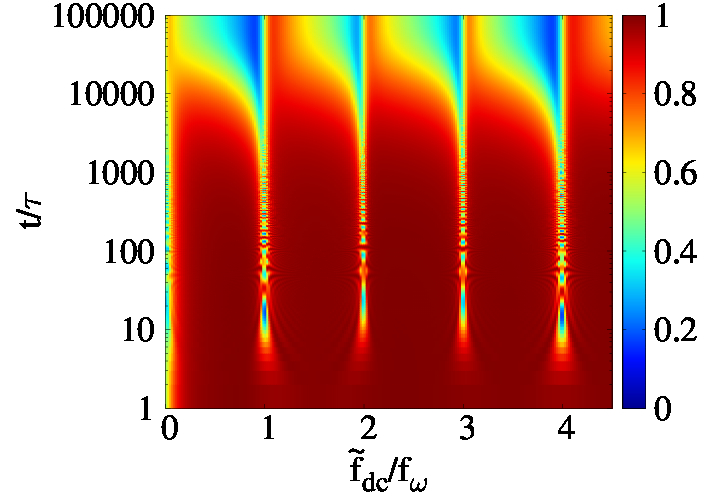}
		\caption {}
		\label{fig:3a}
	\end{subfigure}
	\begin{subfigure}[ht!]{0.45\textwidth} 
	\centering
		\includegraphics[width=0.85\linewidth,clip]{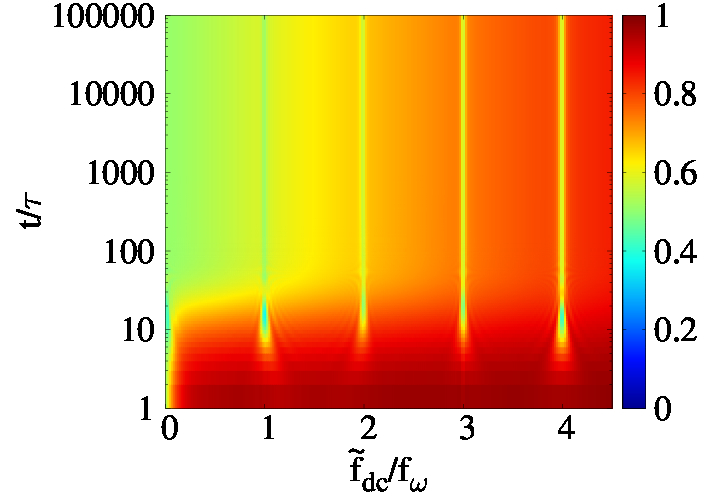}
		\caption {}
		\label{fig:3b}
	\end{subfigure}
	\caption{(color online)  Intensity plots of the population $P_+$ 
		as a function of  ${\tilde f}_{dc}$ and driving time $t$. (a) Longitudinal coupling for 
		$\gamma^{(f)}=0.001$. (b) Transverse coupling  for  $\gamma^{(ch)}=0.001$. See text for details.}
	\label{fig:3}
\end{figure}

For the longitudinal coupling case we find in  Fig.\ref{fig:2} that
the equality $t_d=2t_r$ is satisfied at the multiphoton resonances.  Thus at the resonances
the dephasing mechanism vanishes, similarly to what is usually found 
for the static case at the ``sweet spot'' $\tilde{f}=0$ \cite{fei_2016,bylander_2011}. Away from resonances
is $t_d\ll t_r$, showing a large time scale separation between
decoherence and relaxation, due to strong dephasing. 
We have also obtained an analytic expression for the rates $\Gamma_r=1/t_r$ and the decoherence rate $\Gamma_d=1/t_d$ employing  a RWA approximation for detunings  near the $n$-photon resonance, $\tilde{f}\sim nf_\omega$, which are  in good agreement with  these numerical results (see the Appendix for a detailed calculation).
In the case of longitudinal noise, the relaxation rates can be estimated as:
\begin{eqnarray*}
	\Gamma_{_r} & = & |\lambda_f\sin(2\varphi)|^{2}[g(-\Omega_{n})+
	g(\Omega_{n})],\\
	\Gamma_{d} & = & \frac{\Gamma_r}{2}+|\lambda_f\cos(2\varphi)|^{2}g(0),
\end{eqnarray*}
 with $\cos(2\varphi)=\epsilon_{n}/\varOmega_{n}$, $\sin(2\varphi)=\Delta_{-n}/\varOmega_{n}$,
$\epsilon_{n}=4\pi I_p(\tilde{f}-nf_\omega$). The generalized Rabi frequency is
$\varOmega_{n}=\sqrt{\epsilon_{n}^{2}+\Delta_{-n}^{2}}$ and $\Delta_{-n}=\Delta J_{-n}(x)$ with $x \equiv {f_{ac}/f_{\omega}}$.
At the resonance is $\cos(2\varphi)=0$ and $\sin(2\varphi)=1$, thus $\Gamma_d=\Gamma_r/2$ and $\Gamma_r$ is maximum. 
Away from resonance the dephasing rate is maximum and $\Gamma_\phi=\Gamma_d-\Gamma_r/2\sim \lambda_f^2 g(0)\approx \lambda_f^2\gamma kT$ (assuming $\cos(2\varphi)\sim 1$).
This in agreement with the exact numerical results of Figures \ref{fig:2}(a) and (b) 
where $t_d\ll t_{exp}\ll t_r$ away from resonance for the longitudinal case. 
This  time scale separation  allows the dynamic transition described in Ref.[\onlinecite{ferron_2016}] and  is also shown
in Fig.\ref{fig:3}(a). We see that while
$P_+$ remains symmetric  around a resonance for  $t_d<t<t_r$,  there is a dynamic transition to the  antisymmetric behavior for $t>t_r$.  

On the other hand, for the transverse coupling case we find in  Figs.\ref{fig:2}(a) and (b) that the equality $t_d=2t_r$ is  reached out of resonance, i.e. opposite to the longitudinal case, while near the resonances the (small) dephasing gives $t_d\lesssim 2t_r$. The RWA calculation detailed in the Appendix is also consistent with this numerical result. For  the transverse coupling  we got:
\begin{eqnarray*}
	\Gamma_{r} & = & |\lambda_{ch}J_{-n}(x)\cos(2\varphi)|^{2}[g(-\Omega_{n})+g(\Omega_{n})] ,\\ \Gamma_{d} & = & \frac{\Gamma_r}{2}+|\lambda_{ch}J_{-n}(x)\sin (2\varphi)|^{2}g(0).
\end{eqnarray*}
In this case, away from resonance is $|\epsilon_n|\gg\Delta_{-n}$ implying $\sin(2\varphi)\sim0$, $\cos(2\varphi)\sim1$  and thus $\Gamma_d\approx\Gamma_r/2$. At a  resonance the opposite condition is satisfied: the dephasing rate is maximum and thus  $\Gamma_\phi \propto|\lambda_{ch}|^2 g(0)\approx |\lambda_{ch}|^2\gamma kT$.

In addition, for transverse coupling the system tends to relax fast to the steady state in comparison  to the longitudinal coupling case (assuming the same  coupling strengths $\lambda 's$).
Note  that, in the RWA calculation,  out of resonance is $\cos2\varphi^2 \gg \sin2\varphi^2 $ and then
$\Gamma_r^{\rm transverse}\gg\Gamma_r^{\rm longitudinal}$ for the same  $\lambda$.
This relatively fast relaxation  is evident in Fig.\ref{fig:3}(b) where the steady state is quickly reached and no symmetry change  around the resonance is observed.

\subsection{Mixed noise}
\label{subsec:IIB}

We deal now with the more general case when two  sources of independent noise are taken into account, as formulated in  Sec.\ref{sec:I}, and we consider the  two system-bath couplings with  ${\mathcal{A}}^{f}=-\lambda_f {\sigma}_z$
and   ${\mathcal{A}}^{ch}=-\lambda_{ch}{\sigma}_y$.
For simplicity we consider as before  $J^{(f)}(\omega)=J^{(ch)}(\omega)=\gamma \omega e^{-\omega/\omega_c}$. 
 
\begin{figure}[htb!]
	\begin{subfigure}[ht!]{0.45\textwidth} 
	\centering
		\includegraphics[width=0.85\linewidth]{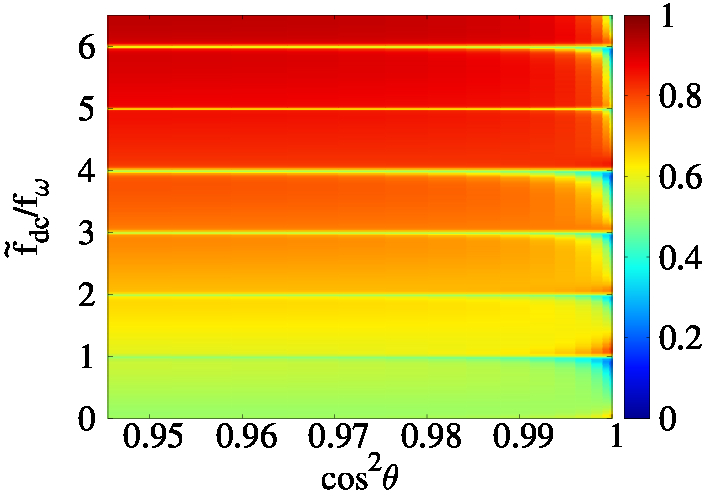}
		\caption {}
		\label{fig:4a}
	\end{subfigure}
	\begin{subfigure}[ht!]{0.45\textwidth} 
	\centering
		\includegraphics[width=0.85\linewidth]{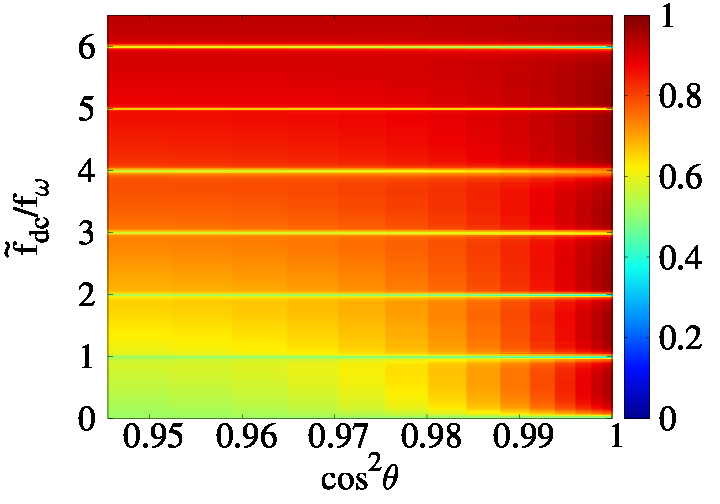}
		\caption {}
		\label{fig:4b}
	\end{subfigure}
	\begin{subfigure}[ht!]{0.45\textwidth} 
	\centering
		\includegraphics[width=0.85\linewidth]{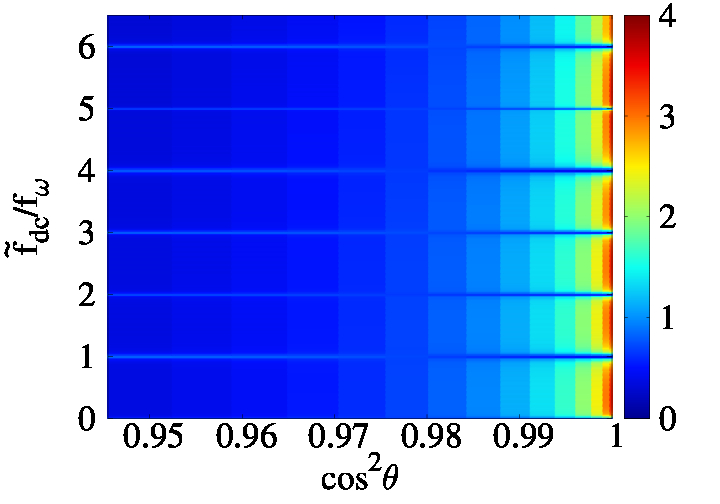}
		\caption {}
		\label{fig:4c}
	\end{subfigure}
	\caption{ Intensity plots of the population $P_+$ 
		as a function of  ${\tilde f}_{dc}$ and the mixing parameter $\cos^2\theta$. $\gamma^{(f)}=\gamma\cos^2\theta$ and
		$\gamma^{(ch)}=\gamma\sin^2\theta$, with $\gamma=0.001$.
		(a) $t=\infty$, (b) $t=1000\tau$, (c) Intensity plot of the ratio of  $\log({2t_r/t_{d}})$.}
	\label{fig:4}
\end{figure}

In order to compare the relative coupling strengths we define $\lambda_f=\cos\theta$
 and $\lambda_{ch}=\sin\theta$.  We     plot in Fig.\ref{fig:4}   $P_+$ for $f_{ac}=0.003$, as a function of  the mixing parameter $\cos^2\theta$ and ${\tilde f}_{dc}$, for the stationary case (Fig.\ref{fig:4}(a)) and for finite time $t=1000\tau$ (Fig.\ref{fig:4}(b)).
In both cases  the plots exhibit a  behavior  similar to the one obtained for the transverse coupling  (see Fig.\ref{fig:3b}), for almost all the range of  $\cos^2\theta$. Only when $(\lambda^{(ch)}/\lambda^{(f)})^2<0.005$ the  typical features of the pure longitudinal case (already described) are observed.

In agreement with the observed response in  $P_+$, Fig.\ref{fig:4}(c)  shows that $2t_r/t_d\sim 1$ for almost all the range
of the mixing parameter, and only when approaching the longitudinal case, $(\lambda^{(ch)}/\lambda^{(f)})^2<0.005$, the time scale separation  $2t_r/t_d \gg 1$ is observed in the off-resonant regions. 



\section{Multilevel regime: LZS diamonds}
\label{subsec:IIC}

The previous  analysis can be extended to  the multilevel regime which  corresponds to realistic parameters of the FQ. 
We  focus on the dynamics of the four  lowest energy levels of the device, where the spectrum shows a rich structure of avoided crossings as a function of the dc detuning \cite{oliver_2005,ferron2_2010}. We solve the Floquet-Markov equations for the Hamiltonian of Eq.(\ref{eq:1}), restricted to the subspace of spanned by the four lowest levels. Here we will compare the LZS patterns  for  pure flux noise and pure charge noise. In  both cases we consider  an Ohmic  bath with spectral density $J(\omega)=\gamma\omega e^{-\omega/\omega_c}$ at temperature $T$, but different coupling operators.
For pure flux (longitudinal) noise we take 
$${\mathcal{A}}^{(flux)}=2\pi\alpha\sin (2\pi f + 2\varphi_{m}),$$
which in the case of the two lowest levels subspace corresponds to ${\mathcal{A}}\approx-\lambda_f{\sigma}_{z}$, with
$\lambda_f=2\pi\alpha |\langle+|\sin (2\pi  f + 2{\varphi}_{m})|+\rangle|\approx 4.5$, for FQ parameters $\alpha=0.8$ and $\eta=0.25$.

In the case of charge (transverse) noise the system operator 
is 
$${\mathcal{A}}^{(charge)}={ n}_{m}=i\frac{\eta^2}{2(1+2\alpha)}\frac{\partial}{\partial\varphi_{m}},$$
which for the two lowest levels subspace gives ${\mathcal{A}}\approx-\lambda_{ch}{\sigma}_{y}$, with $\lambda_{ch}=\frac{\eta^2}{4+8\alpha}|\langle-|{ n}_m|+\rangle|\approx 3 \times 10^{-4}$
for the same FQ parameters. 

Notice that, after introducing realistic parameters, we obtain
$\lambda_{ch}\ll\lambda_{f}$. For this parameter values, since $(\lambda^{(ch)}/\lambda^{(f)})^2\sim 10^{-8} \ll 0.005$, it is irrelevant to study the mixed dynamics with both types of couplings since the transverse noise effects will be unobserved. Thus, we will consider only the cases of pure longitudinal and pure transverse noise in this section to analyze the effect of each type of noise on the LZS patterns separately.
\begin{figure}[!htb]
	
	\begin{subfigure}[ht!]{0.45\textwidth} 
	\centering
		\includegraphics[width=0.85\linewidth]{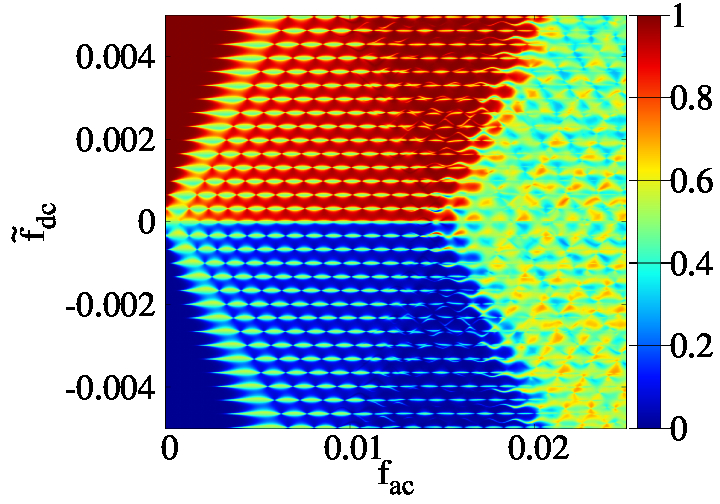}
		\caption {}
		\label{fig:5a}
	\end{subfigure}
	\begin{subfigure}[ht!]{0.45\textwidth} 
	\centering
		\includegraphics[width=0.85\linewidth]{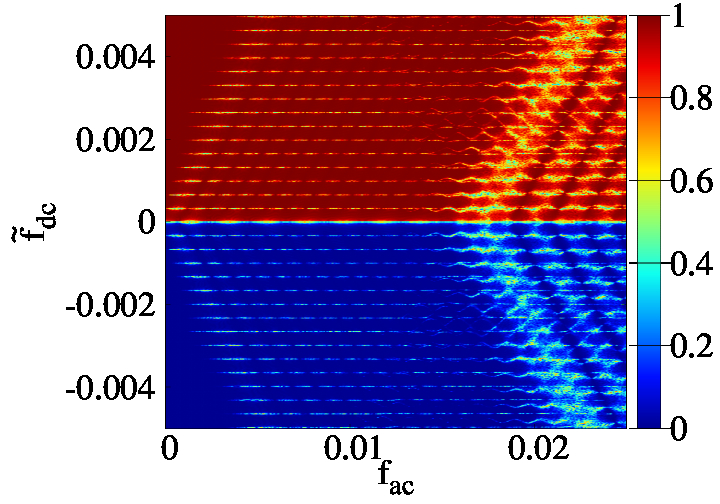}
		\caption {}
		\label{fig:5b}
	\end{subfigure}
	\caption{LZS interference patterns. Plots of $P_+$ as a function of the driving amplitude $f_{ac}$ and dc detuning ${\tilde f}_{dc}$ for $t=1000 \tau$. (a) Flux noise. (b) Charge noise. The calculations were performed for $\omega_{0}= 2\pi/\tau= 0.003 E_J/\hbar$, Ohmic bath  at $T=0.0014 E_J/k_B\sim 20$mK  and $\gamma=0.001$ (see text for details).
}
	\label{fig:5}
\end{figure}
\begin{figure}[!htb]
	
	\begin{subfigure}[ht!]{0.45\textwidth} 
	\centering
		\includegraphics[width=0.85\linewidth]{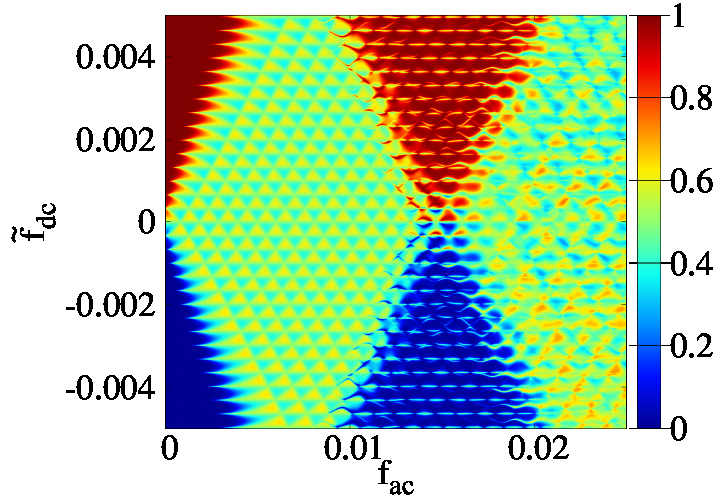}
		\caption {}
		\label{fig:6a}
	\end{subfigure}
	\begin{subfigure}[ht!]{0.45\textwidth} 
	\centering
		\includegraphics[width=0.85\linewidth]{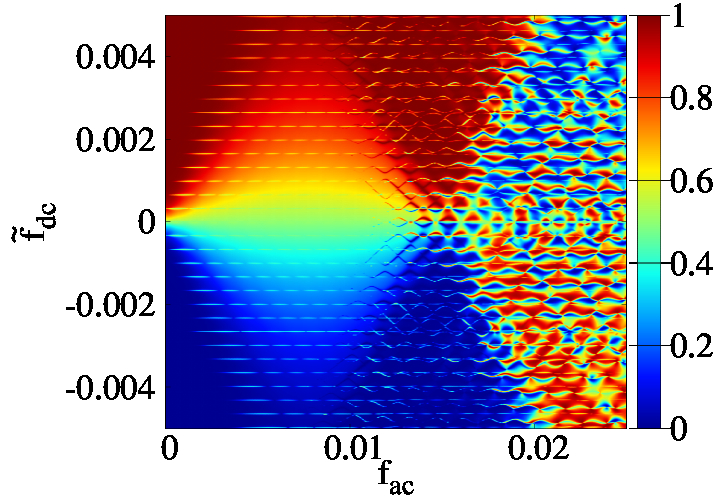}
		\caption {}
		\label{fig:6b}
	\end{subfigure}
	\caption{LZS interference patterns. Plots of $P_+$ as a function of the driving amplitude $f_{ac}$
and dc detuning ${\tilde f}_{dc}$ for the asymptotic regime, $t\rightarrow\infty$. (a) Flux noise. (b) Charge noise.
Same parameters as in Fig.(\ref{fig:5}).
}
\label{fig:6}
\end{figure} 
In Fig.\ref{fig:5}  we plot $P_+$ as a function of the driving amplitude $f_{ac}$ and dc detuning ${\tilde f}_{dc}$, for  $t_{exp}=1000\tau$ and in Fig.\ref{fig:6} for the steady state. The  LZS  interference patterns show the 
typical "diamonds'' structure  for increasing $f_{ac}$, concomitant with  the  additional transitions at the avoided crossings between different  energy levels \cite{berns_2008, ferron_2012, ferron_2016}.  
We plot a range of $f_{ac}$ that shows the first LZS diamond, D1, and the lower half 
of the second LZS diamond, D2.
D1 can be described in terms of  the dynamics of the 
two lowest energy levels, the region between D1 and D2
involves the dynamics of the three lowest energy levels;
while D2 includes the four lowest
energy levels (see Ref.[\onlinecite{berns_2008}] for a complete description of  the multilevel
LZS diamonds).

For finite time $t\sim t_{exp}$, symmetric resonance lobes are observed within D1 for both
types of coulpling. However, for the transverse coupling case 
(Fig.\ref{fig:5}(b)) the resonance lobes are narrower  than for the longitudinal coupling 
(Fig.\ref{fig:5}(a)). The width of the resonance peaks is roughly proportional to the decoherence rate $1/t_d$ \cite{shevchenko_2010,oliver_2005}. As analyzed in the previous section, in the transverse case dephasing mechanisms vanish out of resonance and $1/t_d$ is minimum. On the other hand, the dephasing rate grows out of resonance in the longitudinal case, and $1/t_d$ is large.

In the steady state the
differences among the two types of coupling are stronger. While  for the  longitudinal coupling, Fig.\ref{fig:6}(a) shows  the triangular checkerboard pattern characteristic of antisymmetric resonances together with population inversion (both features described in detail in Refs.\cite{ferron_2012,ferron_2016}), for the transverse coupling (Fig.\ref{fig:6}(b)), D1  exhibits a predominant   background  
with a  symmetric  lobe in $P_+$ around $\tilde{f_{dc}}=0$.
Within D2 and for the longitudinal coupling case, the patterns look qualitatively similar at  finite time $t\sim t_{exp}$ and in the steady state, respectively. On the other hand, for the transverse coupling case, the steady state profile  shows a strong population inversion in D2, absent at  finite time $t\sim t_{exp}$.
\begin{figure}[htb!]
	\centering
	\includegraphics[width=0.93\linewidth]{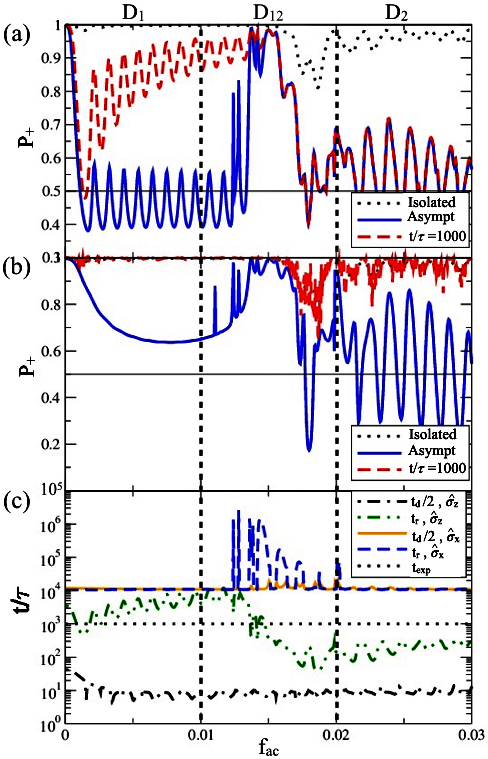}	\caption{$P_+$ vs $f_{ac}$ for  $\tilde{f}_{dc}\equiv 2.7 f_\omega$, for  $t=1000 \tau$ (red dashed line), asymptotic state, $t\rightarrow{\infty}$,  (blue solid line) and for the isolated case (dots). (a) Flux noise. (b)  Charge noise. (c)  $t_r/\tau$ for flux noise (dashed-double dotted line) and charge noise (dashed line). ${t_d}/2\tau$ for flux noise (dot-solid line) and charge noise (solid line). $t_{exp}= 1000\tau$ (dotted line). Vertical  dashed lines are guides for the eyes to  show the boundaries of  diamonds D1, D2 and the region in between, named D12, for the  value of $\tilde{f}_{dc}\equiv 2.7 f_\omega$.}
\label{fig:7}
\end{figure}

To understand the  different time scales, we  plot $P_+$ at a finite time and in the steady state, as a function of the driving amplitude
$f_{ac}$ for a fixed off-resonant value of 
detuning $\tilde{f}_{dc}=0.0009\equiv 2.7 f_\omega$, for the the longitudinal coupling (Fig.\ref{fig:7}(a))  and  for the transverse coupling (Fig.\ref{fig:7}(b)),  respectively.  The time scales  of decoherence and relaxation, $t_d$ and $t_r$, are plotted in Fig.\ref{fig:7}(c).
In the previous section we concluded that for same coupling strengths, 
$\lambda_{ch}=\lambda_{f}$, the transverse coupling leads to a faster relaxation rate.
Here, the smallness of $\lambda_{ch}$ gives a much larger $t_r$ than in the $ \lambda_{ch}\sim 1$ case analyzed previously.  It is interesting to note
in Fig.\ref{fig:7}(c) that the resulting $t_r$ for the transverse coupling turns out to be of the same order or larger than in the  longitudinal case.
 
From Fig.\ref{fig:7}(c) it  follows that for the longitudinal case and within D1,  there is a large time scale separation  $t_d\ll t_{exp}\ll t_r$, in agreement with the different behaviors of  $P_+(t_{exp}=1000\tau)$ and $P_+(\infty)$ seen in  Fig.\ref{fig:7}(a). The relaxation time strongly depends on $f_{ac}$ and within D2, $t_r$ is reduced two orders of magnitude, leading to $t_r<t_{exp}$ and therefore $P_+(t_{exp})\approx P_+(\infty)$.

For  the transverse coupling case, see Fig.\ref{fig:7}(b), 
within both diamonds D1 and D2 the time scales $t_r$ and $t_d$  are both larger than $t_{exp}=1000\tau$ and  nearly independent of $f_{ac}$ (except in the transition between D1 and D2). Thus,  for this type of coupling the steady state behavior could not be seen at the experimental time scale neither for D1 nor for D2. Furthermore, it is also evident that the decoherence rate is minimum since $t_r\sim t_d/2$ in all the range of $f_{ac}$, even beyond the two level regime discussed in the previous section.

From our analysis it is clear that the experimental results of Refs.[\cite{oliver_2005,oliver_2009}]  do not correspond to any of the steady state LSZ patterns of Fig.\ref{fig:6}, since these experiments do not show neither the anstisymmetric resonance patterns of the longitudinal coupling nor the background lobe for off-resonant population of the transverse coupling. In addition, the extremely narrow resonance lobes of Fig.\ref{fig:5}(b) for the transverse coupling do not seem to represent well the experimental data. 
The symmetric resonance lobes of the experimental LZS patterns are more in agreement with the case of Fig.\ref{fig:5}(a) for longitudinal coupling. This conclusion is consistent with the noise spectroscopy measurements of Refs.\cite{bylander_2011,yan_2013,yoshihara_2014} that found  that the transverse noise is very small for FQ devices.


\section{Concluding Remarks}

We have performed a systematic analysis of environmental noise effects for  a strongly driven FQ device, considering a realistic {\it multilevel dynamics} and emphasizing the behavior at {\it different time scales}. 

A main outcome of our work is to expose  the LZS interferometry  as a tool to unveil the  type of system-bath coupling, where the presence of symmetric (asymmetric)  n-photon resonances in the stationary patterns reveals the  nature of the noise, i.e. transverse (longitudinal) system-bath coupling.

In addition the analysis of the relaxation and decoherence time scales  shows  that the ratio $t_{r}/2t_{d}$ 
is also  extremely sensitive to  the  type of system-bath coupling and  might change significantly when a  n-photon resonance is tuned.



For time scales prior to relaxation,  the LZS interferometric patterns  also exhibit two well differentiated behaviours depending on  the noise sources. 
Along this line, our results  for the FQ device in the regime of strong driving (beyond the TLS regime)  shed light on the interpretation  of  the  experimental LZS diamonds obtained  in  Ref.\cite{berns_2008,oliver_2009}  for a driven FQ with long relaxation times. The symmetric resonances lobes observed in  Ref.\cite{berns_2008,oliver_2009}  are in agreement with the longitudinal noise scenario shown in Fig.\ref{fig:5}(a). However,   to conclusively discard other possible scenarios, 
experiments should be performed for larger driving times, in order to reach the steady state after full relaxation with the bath degrees of freedom.

Experimental studies of noise spectroscopy for the FQ, when driven at the first resonance, have shown that flux noise is the dominant  source of decoherence\cite{yan_2013,yoshihara_2014}. This result is also consistent with the scenario of longitudinal noise found  in Fig.\ref{fig:5}(a) for the case  of multiphoton resonances and large amplitudes. 
However, flux noise power spectrum at low frequencies 
has shown   $1/f$ behavior \cite{bylander_2011,yan_2013,yoshihara_2014}. Thus, to better account  noise effects  in the steady state or  long time limit, future studies based on a non-markovian description \cite{stace_2013} would be interesting.


Even when we have considered specific parameters of the FQ, our results can be also useful for other qubits and artificial atoms devices, in which the amplitude spectroscopy technique based on LZS interferometry has been implemented during the last years \cite{sillanpaa_2006,mark_2007,wilson_2010, petta_2010, kling_2010,huang_2011}.

We acknowledge  financial support from CNEA, CONICET (PIP11220150100756), UNCuyo (P 06/C455) and ANPCyT (PICT2014-1382, PICT2016-0791).

\appendix
\renewcommand\thefigure{\thesection.\arabic{figure}} 

\section{The rotating wave approximation: dressed basis}
\label{ap:A}



In this section we briefly revisit the Rotating Wave Approximation (RWA) applied to multiphoton resonances \cite{wilson_2007,ashhab_2007,shevchenko_2010,kmetic_1986}. 
We start by considering the general Two Level System  (TLS) Hamiltonian:
\begin{equation}
{\mathcal{H}}_{TLS}(t)=-\frac{\epsilon(t)}{2}{\sigma}_{z}-\frac{\Delta}{2}{\sigma}_{x}
\label{eq:A1}
\end{equation}
where $\epsilon(t)=\epsilon_{0}+A\cos(\omega_{0} t)$. The parameter $\epsilon_{0}$ is the polarization energy of the qubit,   $A$ and $\omega$  the amplitude and frequency of the driving, respectively. By applying  the unitary transformation
$|\tilde{\psi}(t)\rangle= U_{\phi}(t)|\psi(t)\rangle$, with $ U_{\phi}(t)=e^{-i\frac{\phi (t)}{2}{\sigma}_{z}}$ and $\phi(t)= \int \epsilon(t) \,dt = \epsilon_{0}t + A/\omega_{0} \sin(\omega_{0} t)$, the transformed Hamiltonian reads:
\begin{equation}
{\tilde{H}}=-\frac{(\epsilon-\dot{\phi} (t) )}{2}{\sigma}_{z}-\frac{\Delta}{2}(e^{-i\phi (t) }{\sigma}_{+}+h.c).
\label{eq:A2}
\end{equation} 
Replacing $\phi(t) \rightarrow \phi_{n}(t)=n\omega_{0} t+\frac{A}{\omega_{0}}\sin\omega_{0} t$ (which is equivalent to
take the resonance condition  $\epsilon_{0} \sim n\omega_{0}$)  the Hamiltonian transforms to:
\begin{equation}
{\tilde{H}}=-\frac{(\epsilon_{0}-n\omega_{0})}{2}{\sigma}_{z}-\frac{\Delta}{2}(e^{-in\omega_{0} t}e^{-i\frac{A}{\omega_{0}}\sin\omega_{0} t}{\sigma}_{+}+h.c).
\label{eq:A3}
\end{equation} 
Using in addition that $e^{ix\sin a}=\sum_{k=-\infty}^{k=+\infty}J_{k}(x)e^{ika}$,  with $J_{k}(x)$ the Bessel function of order $k$, we can write:
\begin{equation}
e^{in\omega_{0} t}e^{i\frac{A}{\omega_{0}}\sin\omega_{0} t}=\sum_{k=-\infty}^{k=+\infty}J_{k}(\frac{A}{\omega_{0}})e^{i(k+n)\omega_{0} t}\simeq J_{-n}(\frac{A}{\omega_{0}})
\label{eq:A4}
\end{equation}
where in the last step we have performed a rotating wave approximation (RWA), for $|\epsilon_{0}-n\omega_{0}|\ll\varDelta$. In this way, we  finally obtain the  TLS Hamiltonian written in the RWA as: 
\begin{equation}
{\tilde{H}}\simeq {\tilde{H}}_{n} =-\frac{(\epsilon_{0}-n\omega_{0})}{2}{\sigma}_{z}-\frac{\Delta_{-n}}{2}{\sigma}_{x},
\label{eq:A5}
\end{equation}
with $\Delta_{-n}=\Delta J_{-n}(\frac{A}{\omega_{0}})$. 

Notice that after the RWA we have obtained an effective time-independent ``dressed" Hamiltonian.
Going a  step further, we proceed to diagonalize  ${\tilde{H}}_{n} $ considering the operator $U_{r}=\cos (\varphi){\sigma}_{z}+\sin (\varphi){\sigma}_{x}$. After applying such transformation, we finally obtain the ``dressed" Hamiltonian
\begin{equation}
 {{H}}_{r}=U_{r}{\tilde{H}}_{n}  U_{r}^{-1}=-\frac{\varOmega_{n}}{2}{\sigma}_{z},
\label{eq:A6}
\end{equation}
with $\cos(2\varphi)=\epsilon_{n}/\varOmega_{n}$, $\sin(2\varphi)=\Delta_{-n}/\varOmega_{n}$,
$\epsilon_{n}=\epsilon_{0}-n\omega_{0}$, and $\varOmega_{n}=\sqrt{(\epsilon_{0}-n\omega_{0})^{2}+\Delta_{-n}^{2}}$  the generalized Rabi frequency.  It is worth noting that the eigenenergies of ${H}_{r}$ exhibit an  avoided crossing with an effective ``dressed" gap $\Delta_{-n}$, and  the associated  eigenstates form the ``dressed" basis.

\section{Calculation of relaxation and decoherence rates in the rotating wave approximation}
\label{ap:B}

The  dynamics of an open system can be described by the total Hamiltonian:
\begin{equation}
 H_{T}(t)={\mathcal{H}}_{TLS}(t)+ H_{B}+{H}_{int},
\label{eq:B1}
\end{equation}
where ${\mathcal{H}}_{TLS}(t)$ is  the driven TLS Hamiltonian, $ H_{B}$ the bath term and 
\begin{equation}
{H}_{int}={\mathcal{A}}\otimes {\mathcal{B}},
\label{eq:B2}
\end{equation} the system-bath interaction term. In the present analysis the system operator, ${\mathcal{A}}$,  can be  ${\mathcal{A}}_{z}=\lambda_z{\sigma}_{z}$ or ${\mathcal{A}}_{x}=\lambda_x{\sigma}_{x}$ and  
${\mathcal{B}}$ is the bath operator. 

The von Neumann equation for  time-evolution of the system  described by the total Hamiltonian (\ref{eq:B1}) is ((taking $\hbar=1$)
\begin{equation}
\frac{\partial   \rho_{T}(t)}{\partial t}=-i[ H_{T}(t), \rho_{T}(t)],
\label{eq:B6}
\end{equation} with $ \rho_{T}(t)$ the density matrix of the global system.

We start by  defining  $ H_{0}(t)= {\mathcal{H}}_{TLS}(t)+  H_{B}$, and the associated  evolution operator $ U_{0}(t)=\hat{\mathcal{T}}e^{-i\int  H_{0}(t)\,dt}$ . Therefore, in the Interaction Picture  the transformed operators  are $ { \tilde{\rho} }(t)  =   U_{0}^{\dagger}(t)  \rho (t) U_{0}(t)$ and ${ \tilde{H} }_{int}(t) =   U_{0}^{\dagger}(t){H}_{int} U_{0}(t)$, and
Eq.(\ref{eq:B6})  reads:
\begin{equation}
\frac{\partial   { \tilde{\rho_{T}} } (t)}{\partial t}=-i[{ \tilde{H}} _{int}(t),  { \tilde{\rho_{T}} } (t)].
\label{eq:B7}
\end{equation} 

After defining the system reduced  density matrix $ \rho=Tr_{B}( \rho_{T})$ and performing  the Born-Markov approximation, we get:
\begin{eqnarray}
\frac{\partial  { \tilde{\rho} }  }{\partial t} & = & -\int_{0}^{\infty}dt'\{G(t')[{\tilde{ \mathcal{A}}}(t){\tilde{ \mathcal{A}}}(t-t'){ \tilde{\rho} }-{\tilde{ \mathcal{A}}}(t-t'){ \tilde{\rho} }{\tilde{ \mathcal{A}}}(t)]\nonumber \\
&  & +G(-t')[{ \tilde{\rho} }{\tilde{ \mathcal{A}}}(t-t'){\tilde{ \mathcal{A}}}(t)-{\tilde{ \mathcal{A}}}(t){ \tilde{\rho} }{\tilde{ \mathcal{A}}}(t-t')],
\label{eq:B8}
\end{eqnarray} with $G(t)=\mathrm{Tr}_{B}\Big({\mathcal{B}}(t) {\mathcal{B}}(0)  \rho_{B}\Big)$ the bath correlation function and ${\tilde{ \mathcal{A}}}(t)= U_{0}^{\dagger}(t) {\mathcal{A}}  U_{0}(t)$. 

Now, we set the decomposition
\begin{equation}
{\tilde{ \mathcal{A}}}(t)=\sum_{\omega}e^{-i\omega t} P_{\omega}=\sum_{\omega}e^{i\omega t} P_{\omega}^{\dagger}=\sum_{\omega>0}e^{-i\omega t} P_{\omega}+e^{i\omega t} P_{\omega}^{\dagger}
\label{eq:B9}
\end{equation} 
with $ P_{-\omega} =  P_{\omega}^{\dag} $. Moreover, we can define  $\Gamma(\omega)=\int_{0}^{\infty}dte^{-i\omega t}G(t)=\frac{1}{2}g(\omega)+is(\omega)$\cite{breuer_petruccione_2006}.

After performing the secular approximation,  Eq.(\ref{eq:B8}) can be expressed in the Lindblad form as:
\begin{equation}
\begin{aligned}
\frac{\partial{ \tilde{\rho} }}{\partial t}   = -i[ H_{L},{ \tilde{\rho} }]+\sum_{\omega}g(\omega)( P_\omega{ \tilde{\rho} } P_{\omega}^{\dagger}-\frac{1}{2}\{ P_{\omega}^{\dagger} P_{\omega},{ \tilde{\rho} }\})
\end{aligned}
\label{eq:B10}
\end{equation} with the Hamiltonian
\begin{equation}
 H_{L}=\sum_{\omega}s(\omega) P_{\omega}^{\dagger} P_{\omega}.
\label{eq:B11}
\end{equation}


We now proceed to transform the system operator  ${\mathcal{A}}$ into the ``dressed'' representation \cite{wilson_2007}, ${\mathcal{A}}\rightarrow {\mathcal{A}}_{r}$. 
Following the procedure described previously, we  perform  the transformation ${\mathcal{A}}_{r}(t)= U_{r} U_{n}{\mathcal{A}}U_{n}^{-1} U_{r}^{-1}$, with $U_{r}=\cos (\varphi){\sigma}_{z}+\sin (\varphi){\sigma}_{x}$ and $ U_{n} (t)= e^{-i  \frac{1}{2}\phi_{n}(t){\sigma}_{z}} $.  
For a system operator of the form
${\mathcal{A}}= \lambda( \cos\theta {\sigma}_{z}+ \sin\theta {\sigma}_{x})$, we obtain the
transformed ${\mathcal A}_r$ as: 
\begin{equation}
\begin{aligned}
	{\mathcal{A}}_{r}(t)=a_{x}(t){\sigma}_{x}+a_{y}(t){\sigma}_{y}+a_{z}(t){\sigma}_{z},
\end{aligned}
\label{eq:B13}
\end{equation} with  the coefficients $a_{i}(t)$, $i=x,y,z$, satisfying the following relations:
\begin{eqnarray*}
	a_{x}(t) & = & a_{x0}+a_{xc}\cos\phi_{n}(t)\approx a_{x0}+a_{xc}c_0,\\
	a_{y}(t) & = & a_{ys}\sin\phi_{n}(t) \approx 0,\\
	a_{z}(t) & = & a_{z0}+a_{zc}\cos\phi_{n}(t)\approx a_{z0}+a_{zc}c_0,
\end{eqnarray*} with $a_{x0}=\lambda\cos\theta\sin2\varphi$, $a_{xc}=-\lambda\sin\theta\cos2\varphi$, $a_{ys}= -\lambda\sin\theta$, $a_{z0}=\lambda\cos\theta\cos2\varphi$ and $a_{zc}=\lambda\sin\theta\sin2\varphi$. In the last step we have
performed the RWA as in Eq.(\ref{eq:A4}), 
with $c_{0} = J_{-n}(\frac{A}{\omega_{0}})$.

Transforming ${\mathcal{A}}_{r}$ to the Interaction picture one gets:
\begin{equation}
\begin{aligned}
	{\tilde{ \mathcal{A}}}_{r}(t)=&(a_{x0}+a_{xc}c_{0})e^{i\Omega_{n}t}{\sigma} _{+}+  (a_{x0}+a_{xc}c_{0})e^{-i\Omega_{n}t}{\sigma}_{-}\\
	&+  (a_{z0}+a_{zc}c_{0}) {\sigma}_{z},
	\end{aligned}
	\label{eq:B10}
\end{equation}
To obtain the Linblad equation, we rewrite the above equation in terms of the decomposition 
of Eq.(\ref{eq:B9}),
\begin{equation}
\begin{aligned}
	{\tilde{ \mathcal{A}}}_{r}(t) &= P_{+}({\Omega_{n}})e^{i\Omega_{n}t} + P_{-}(-{\Omega_{n}})e^{-i\Omega_{n}t} + P_{0},\\
	&= P_{+}({\Omega_{n}})e^{i\Omega_{n}t} + P^{\dag}_{+}({\Omega_{n}})e^{-i\Omega_{n}t} + P_{0},\\
	&= \sum_{\omega} P_{\omega} e^{i\omega t},
	\end{aligned}
	\label{eq:11}
\end{equation} with $\omega \in \{0,{\Omega_{n}},-{\Omega_{n}} \}$ and $P_{\omega} = \{ P_{z}(0), P_{+}({\Omega_{n}}) , P_{-}(-{\Omega_{n}}) \} \equiv \{ P_{z}(0), P_{+}({\Omega_{n}}) , P^{\dag}_{+}({\Omega_{n}}) \}$.
The operators $ P_{\omega}$ are:
\begin{equation}
\begin{aligned}
	 P_{z}(0) = & (a_{z0}+a_{zc}c_{0}){\sigma}_{z}=z(0){\sigma}_{0},\\
	 P_{+}({\Omega_{n}}) = & (a_{x0}+a_{xc}c_{0}){\sigma}_{+}= x(\Omega_{n}){\sigma}_{+},\\
	 P_{-}({-\Omega_{n}}) =& P^{\dag}_{+}({\Omega_{n}}) = x(-\Omega_{n}){\sigma}_{-},\\
	\end{aligned}
\label{eq:B12}
\end{equation}  with $z(0)=a_{z0}+a_{zc}c_{0}$ and $x({\Omega_{n}})=a_{x0}+a_{xc}c_{0}$.


Using Eq.(\ref{eq:B12}) in Eq.(\ref{eq:B10}), we obtain the Lindblad equation
\begin{equation}
\begin{aligned}
\frac{\partial{ \tilde{\rho} }}{\partial t}=&-\Gamma_{\uparrow}\left(\begin{array}{cc}
-{ \tilde{\rho} }_{11} & \frac{1}{2} { \tilde{\rho} }_{12}\\
\frac{1}{2}{ \tilde{\rho} }_{21} & { \tilde{\rho} }_{11}
\end{array}\right)-\Gamma_{\downarrow}\left(\begin{array}{cc}
{ \tilde{\rho} }_{22} & \frac{1}{2}{ \tilde{\rho} }_{12}\\
\frac{1}{2}{ \tilde{\rho} }_{21} & -{ \tilde{\rho} }_{22}
\end{array}\right)\\
&-\Gamma_{0}\left(\begin{array}{cc}
0 & { \tilde{\rho} }_{12}\\
{ \tilde{\rho} }_{21} & 0
\end{array}\right)
\end{aligned}
\label{eq:13}
\end{equation}with
\begin{equation}
\begin{aligned}
	\Gamma_{\uparrow} & =  |x(\Omega_{n})|^{2}g(-\Omega_{n})\\
	\Gamma_{\downarrow} & =  |x(\Omega_{n})|^{2}g(\Omega_{n})\\
	\Gamma_{o} & =  |z(0)|^{2}g(0)
\end{aligned}
\label{eq:14}
\end{equation}

After solving  Eq.(\ref{eq:13}), the relaxation $\Gamma_{r}$ and decoherence $\Gamma_{d}$ rates can be computed as 
\begin{equation}
\begin{aligned}
	\Gamma_{d} & =  \Gamma_{\downarrow}+\Gamma_{\uparrow} ,\\
	\Gamma_{r} & =  \Gamma_{0}+\frac{\Gamma_{d}}{2}.
\end{aligned}
\label{eq:15}
\end{equation}

Considering the system-bath coupling term ${\mathcal{A}}_{z} = \lambda_z\sigma_{z}$, the rates in Eq.(\ref{eq:15}) take the form
\begin{equation}
\begin{aligned}
	\Gamma^{z}_{\uparrow} & = |\lambda_z\sin(2\varphi)|^{2}g(-\Omega_{n}),\\
	\Gamma^{z}_{\downarrow} & =  |\lambda_z\sin(2\varphi)|^{2}g(\Omega_{n}),\\
	\Gamma^{z}_{o} & = |\lambda_z\cos(2\varphi)|^{2}g(0).
\end{aligned}
\label{eq:16}
\end{equation} Followed by 
\begin{equation}
\begin{aligned}
	\Gamma^{z}_{d} & =   |\lambda_z\sin(2\varphi)|^{2} \Big( g(-\Omega_{n})  + g(\Omega_{n})\Big) ,\\
	\Gamma^{z}_{r} & = |\lambda_z\cos(2\varphi)|^{2}g(0) + \frac{\Gamma^{z}_{d}}{2}.
\end{aligned}
\label{eq:17}
\end{equation}

For the ${\mathcal{A}}_{x} =\lambda_x \sigma_{x}$ case, the rates are
\begin{equation}
\begin{aligned}
	\Gamma^{x}_{\uparrow} & =  |\lambda_x c_{0}\cos(2\varphi)|^{2}g(-\Omega_{n})\\
	\Gamma^{x}_{\downarrow} & =  |\lambda_x c_{0}\cos(2\varphi)|^{2}g(\Omega_{n})\\
	\Gamma^{x}_{o} & =  |\lambda_x c_{0}\sin(2\varphi)|^{2}g(0).
\end{aligned}
\label{eq:18}
\end{equation} For this case, the calculation of the rates $\Gamma^{x}_{d}$ and  $\Gamma^{x}_{r}$ are rather cumbersome.
We obtain: 
\begin{equation}
\begin{aligned}
	\Gamma^{x}_{d}  =&  |\lambda_x c_{0}\cos(2\varphi)|^{2} \Big( g(-\Omega_{n})  + g(\Omega_{n})\Big)\\
	\Gamma^{x}_{r} =&  |\lambda_x c_{0}\sin(2\varphi)|^{2}g(0).
\end{aligned}
\label{eq:19}
\end{equation} 

In this way, we have extended the calculation of relaxation rates given in the Supplementary Information of \cite{yan_2013} to the case of $n$-photon resonances.

%
%
%
%
%
%
%
%
%
%
%

\bibliography{references}

\begin{thebibliography}{66}
\bibitem [{foo()}]{foot1}%
  \BibitemOpen
  \href@noop {} {\bibinfo  {journal} {For the qubit parameters under
  consideration, one can show that the ratio of matrix elements of the charge
  operators for the lowest energy states is $\frac{|\langle i|{\hat n}_{p}
  |j\rangle|}{|\langle i|{\hat n}_{m} |j\rangle|}\sim 10^{-10}$.}\
  }\BibitemShut {NoStop}%
\bibitem [{com()}]{comment4}%
  \BibitemOpen
  \href@noop {} {\bibinfo  {journal} {Notice that when the different noise
  sources are fully correlated, such that ${\cal B}^{\nu}(t)=a_\nu{\cal B}(t)$,
  it is straightforward to show (after redefining ${\cal A}= \sum_{\nu}
  a_\nu{\cal A}^{\nu}$) that ${H}_{int}={\cal A}\otimes{\cal B}$, i.e. there is
  an effective single noise source}\ }\BibitemShut {NoStop}%
\end{thebibliography}

\end{document}